\documentclass{pasj01}
\usepackage{pdflscape}

\begin{document}
\SetRunningHead{H. Nakanishi and Y. Sofue}{3-D Distributions of {the Total Neutral Gas} in the Milky Way Galaxy}
\Received{2015/08/17}
\Accepted{2015/10/06}

\title{Three-Dimensional Distribution of the ISM in the Milky Way Galaxy: III. {The Total Neutral Gas Disk}}
\author{Hiroyuki \textsc{Nakanishi}\altaffilmark{1,2} and Yoshiaki \textsc{Sofue}\altaffilmark{3}}%
\altaffiltext{1}{Graduate Schools of Science and Engineering, Kagoshima university, 1-21-35 Korimoto Kagoshima}
\altaffiltext{2}{Institute of Space and Astronautical Science, Japan Aerospace Exploration Agency, 3-1-1 Yoshinodai, Chuo-ku, Sagamihara, Kanagawa 252-5210, Japan}
\altaffiltext{3}{Institute of Astronomy, The University of Tokyo, 2-21-1 Osawa, Mitaka,Tokyo 181-0015}
\email{hnakanis@sci.kagoshima-u.ac.jp}
\maketitle

\begin{abstract}
We present newly obtained three-dimensional gaseous maps of the Milky Way Galaxy; HI, H$_2$ and total-gas (HI plus H$_2$) maps, which were derived from the HI and $^{12}$CO($J=1$--0) survey data and rotation curves based on the kinematic distance. 
The HI and H$_2$ face-on maps show that the HI disk is extended to the radius of 15--20 kpc and its outskirt is asymmetric to the Galactic center, while most of the H$_2$ gas is distributed inside the solar circle. The total gas mass within radius 30 kpc amounts to $8.0\times 10^9$ M$_\odot$, 89\% and 11\% of which are HI and H$_2$, {respectively}. The vertical slices show that the outer HI disk is strongly warped and the inner HI and H$_2$ disks are corrugated. 
The total gas map is advantageous to trace spiral structure from the inner to outer disk. Spiral structures such as the Norma-Cygnus, the Perseus, the Sagittarius-Carina, the Scutum-Crux, and the Orion arms are more clearly traced in the total gas map than ever. All the spiral arms are well explained with logarithmic spiral arms with pitch angle of $11\degree$ -- $15\degree$. 
The molecular fraction to the total gas is high near the Galactic center and decreases with the Galactocentric distance. 
The molecular fraction also locally enhanced at the spiral arms compared with the inter-arm regions. 
\end{abstract}


\section{INTRODUCTION}
The interstellar medium (ISM) is one of the major components of the Galaxy and consists of coronal gas, intercloud gas, diffuse clouds, dark clouds, Bok globules, molecular clouds, and H {\sc ii} regions \citep{mye78}. The neutral hydrogen H {\sc i} radiates 1420 MHz line and traces intercloud gas, diffuse clouds, and dark clouds. Though dominant component of molecular clouds is H$_2$ gas, it does not have emission line in the radio wavelength range. Therefore, in general, carbon monoxide molecule CO line is used as a tracer of H$_2$ gas and H$_2$ gas density is calculated with a conversion factor $X_{\rm CO}$. 

All sky HI and CO line survey data in radio wavelength are great tools to reveal the Galactic global structure, because (1) radio signal suffers relatively low extinction due to its long wavelength, (2) neutral hydrogen and molecular gases observed in the radio lines are distributed in the whole Galaxy, and (3) line data provide us with information about distances using a kinematic distance method. 

There have been a lot of efforts to conduct HI and CO line surveys for the Galaxy. \citet{mul57} carried out Kootwijk HI survey with 7.5-m telescope, with which \citet{oor58} constructed the HI distribution map and found spiral structures such as the Sagittarius, the Orion (Local)), and the Perseus arms. Succeeding HI survey works found more spiral structures such as the Cygnus (Outer) arm by analyzing longitude-velocity (LV) diagram \citep{ker69,wea70,ver73}. The latest HI survey conducted with Australia Telescope Compact Array showed that there exist additional spiral structure at the end of the HI disk in the fourth quadrant \citep{mcc04}.
 
\citet{dam87} and \citet{bro88} carried out $^{12}$CO($J=1$--0) line survey for the Galactic plane with 1.2-m telescope, with which the Sagittarius and Carina arms were clearly traced by locating giant molecular clouds in the Galactic plane \citep{coh85,dam86,mye86,gra87}. \citet{san86} carried out CO survey for the first quadrant of the Galaxy with 14-m telescope and traced the Sagittarius and Perseus arms. Using the same CO survey data, \citet{cle88} derived a face-on map of the molecular cloud distribution in the first quadrant. The latest study identified new distant molecular spiral arm in the first quadrant, which is considered to be the continuation of the Scutum–Crux arm \citep{dam11}. 
 
These spiral arms are identified with those in many other tracers such as H {\sc ii} regions \citep{geo76,dow80}, thermal electrons \citep{tay93}, dust emission \citep{dri01}, $^{26}$Al \citep{che96}, synchrotron emission \citep{beu85}, and methanol masers \citep{cas11} besides the HI and CO studies. 
\citet{val05} compiled studies of measurements of spiral arms published from 1980 to early 2005 for the meta-analysis of the Galactic spiral structure and presented spiral arms. Most studies have identified four spiral arms, whose shapes are concluded as logarithmic spiral arms with pitch angle of $\timeform{13D.1} \pm \timeform{0D.6}$ \citep{val15}. 



In Paper I \citep{nak03} and Paper II \citep{nak06a}, we derived HI and H$_2$ gas maps using the latest HI \citep{har97,ker86,bur83} and CO survey data \citep{dam01}, and rotation curves \citep{cle85,deh98}, based on the kinematic distance method. 
In these papers, we showed individual three dimensional maps of HI and H$_2$ disks. The HI gas is vastly distributed in the Galactic disk and it is extended further beyond the edge of the stellar disk. The surface density of the HI gas is larger in the outer Galaxy than in the inner Galaxy. On the other hand, the most of the H$_2$ gas is distributed within the stellar disk. Therefore, the HI data is advantageous to investigate the structure of the outer Galaxy, while CO data is suitable to examine the inner Galaxy. 

In this paper we present total gas maps as well as HI and H$_2$ as the third part of this series. The total gas map is suitable for investigating the entire structure of the Galaxy from the inner to outer part. 
At the same time, we revise the HI map using the newly released LAB survey data \citep{kal05} after Paper I and it has been found that the data analysis method should be improved through the work of Paper II. 

We choose a Cartesian coordinate whose x-axis coincides with the line crossing the Sun and the Galactic center, and whose origin coincides with the Galactic center. The $z$-axis is chosen to be parallel to the rotational axis. We define that the Sun is located at ($-8$ kpc, 0, 0). We also use cylindrical coordinates ($R$, $\theta$, $z$) so that the angle $\theta=180\degree$ coincides with the direction toward the Sun and the angle $\theta=90\degree$ was parallel to $l=90\degree$. 
The Galactic constants $R_0 = 8.0$ kpc (the Galactocentric distance of the Sun) and $V_0 = 217$ km s$^{-1}$ (the solar rotational velocity) are adopted \citep{deh98}.


\section{Data}
HI survey data were taken from the LAB (Leiden-Argentine-Bonn) survey \citep{kal05}, which consists of two survey data sets; Leiden/Dwingeloo Survey conducted with Leiden 25-m telescope whose half-power-beam-width (HPBW) was $\timeform{0D.6}$ covering the northern sky of $\delta \ge -30\degree$ \citep{har97} and Instituto Argentino de Radioastronomia Survey conducted with 30-m telescope whose HPBW was $\timeform{0D.5}$ covering the southern sky of $\delta \le -25\degree$ \citep{baj05}. 
The map grid spacing was $\timeform{0D.5}$ which corresponded to the linear scale of 17 pc at the heliocentric distance of 1 kpc, the LSR (local standard of rest) velocity coverage ranged from $-450$ km s$^{-1}$ to $+400$ km s$^{-1}$ at a resolution of 1.3 km s$^{-1}$, and the rms noise in the brightness temperature was 0.07 -- 0.09 K. Residual systematic errors were reported as below 0.02 -- 0.04 K.
In Paper I, we used only spectra within the latitude range of $|\delta| \le 10\degree$ because the data for the southern hemisphere covered only $|\delta| \le 10\degree$. 
However, we used spectra in the latitude range of $|\delta| \le 30\degree$, which covers more than 99\% of the entire gaseous disk if we assume it is a cylinder with a radius of 20 kpc and thickness of 2 kpc,

CO survey data, which are used to derive H$_2$ gas distribution, were adopted from \citet{dam01}, who compiled CO survey data from \citet{gra87}, \citet{bro89}, \citet{may93}, \citet{bit97} as described in Paper II. The data were obtained with 1.2-m telescopes in the USA and Chile. The HPBW was \timeform{8'.4} -- \timeform{8'.8}, which corresponds to 2.4 -- 2.5 pc at the heliocentric distance of 1 kpc. The map grid spacing was either $1/8\degree$ or $1/4\degree$ depending on a survey area. 
The LSR velocity coverage was 332 km s$^{-1}$ and the velocity resolution was 1.3 km s$^{-1}$. The rms noise in the brightness temperature was 0.12--0.43 K. The Galactic latitude range is $|b|\le \timeform{1D.5}$. Since most of the molecular gas is distributed inside the solar circle and the FWHM (full width of half maximum) of the molecular disk is typically 100 pc, most of the CO gas is covered with spectra in this range. 


The rotation curve data are also the same as those we used in Papers I and II. 
The outer rotation curve ($R > R_0$) was taken from \citet{deh98}, who presented model rotation curves fitted to the observation data available. The outer rotation curve we adopted is the best fitted model among them. The Galactic constant $R_0 = 8.0$ kpc and $V_0 = 217$ km s$^{-1}$ are adopted in this model. 
The inner rotation curve ($R < R_0$) was taken from \citet{cle85}, which was derived from the terminal velocity traced with the Massachusetts-Stoney Brook Galactic equator CO survey data. Though the original rotation curve was fitted with a polynomial function adopting Galactic constants of $(R_0, V_0) =$ (8.5 kpc, 220 km s$^{-1}$) and (10 kpc, 250 km s$^{-1}$), we modified the constants of the polynomial function for the case of $(R_0, V_0) =$ (8.0 kpc, 217 km s$^{-1}$) in order to connect smoothly to the outer rotation as described in Paper I.


\section{Derivations of HI, H$_2$, and Total Gas Maps}
\subsection{Kinematic Distance}
Assuming that the gases circularly rotate around the Galactic center without non-circular motion and that the rotational velocity is constant against the height from the Galactic plane, the heliocentric distance $r$ of HI and CO gases were calculated based on the kinematic distance as follows. 

The LSR velocity $V_{\rm LSR}$ of gas component observed in the direction $(l, b)$ rotating at the Galactocentric distance $R$ is given by the following equation, 
\begin{equation}
V_{\rm LSR} = \left[ {R_0 \over R} V(R) - V_0\right] \sin{l} \cos{b}
\end{equation}
The heliocentric distance $r$ of the gas component can be given by solving the following relation of $r$, $R$ and $R_0$
\begin{equation}
R^2 = r^2 + R_0^2 -2 r R_0 \cos{l}. 
\end{equation}

In the case of the outer Galaxy ($R>R_0$), the heliocentric distance is uniquely determined to be $r =R_0 \cos{l} + \sqrt{R^2-R_0^2 \sin^2{l}}$, while it has two solutions $r =R_0 \cos{l} \pm \sqrt{R^2-R_0^2 \sin^2{l}}$ for a given LSR velocity in the case of the inner Galaxy. However, there is no ambiguity at the tangential point $R=R_0 \sin{l}$, where the LSR velocity reaches the maximum, which is refereed as terminal velocity $V_{\rm t}$, and the heliocentric distance is given by $r =R_0 \cos{l}$.

\subsection{HI and H$_2$ distributions in the outer Galaxy}
In the case of the outer Galaxy, the heliocentric distance can be uniquely determined as explained in the previous subsection.
As described in Papers I and II, volume number densities $n$ [cm$^{-1}$] of HI and H$_2$ gases at the heliocentric distance $r$ in the direction $(l, b)$ are calculated using the following equation, 
\begin{equation}
n = X T_{\rm b} {\Delta V_{\rm r} \over \Delta r}, 
\end{equation}
where $X$ is the conversion factor, for which we adopt $X=1.82\times 10^{18}$ (H cm$^{-2}$ K$^{-1}$ km$^{-1}$ s) for HI and $X=1.8\times 10^{20}$ (H$_2$ cm$^{-2}$ K$^{-1}$ km$^{-1}$ s) for H$_2$ gase \citep{dam01}, respectively. {As described in Paper II, it is known that the conversion factor is likely to increase with Galactocentric distance \citep{ari96}. Therefore, the column density estimated in this paper might be overestimated around the Galactic center and underestimated for the outer Galaxy. H$_2$ map obtained by adopting the conversion factor $X$ taken from \citet{ari96} is shown in Paper II. }

\subsection{HI distribution in the inner Galaxy}
In the case of the inner Galaxy, there is near-far ambiguity that there are two points ($r_1$, $r_2$) giving the same LSR velocity $V_{\rm LSR}$ in a single line-of-sight as explained in section 3.1. In order to solve this problem, we introduced a model of vertical density distribution at $r_i$ \citep{spi42},  
\begin{equation}
n_i(z) = n_{0i} {\rm sech}^2(\ln{(1 + \sqrt{2})}{z - z_{0i} \over z_{1/2}}),  
\end{equation}
where $z_{0i}$ is the height of a midplane from the plane of $b=0\degree$, $n_{0i}$ is volume density at the midplane, and $z_{1/2}$ is half the full width of half maximum (FWHM) of the vertical gas distribution. 
 
Adopting this model, the observed column density $N(b)$ is given as a function of Galactic latitude $b$ with the above model as follows,  
\begin{equation}
N(b) \cos{b} = n_1(r_1 \tan{b})\Delta r_1 + n_2(r_2 \tan{b})\Delta r_2. 
\end{equation} 
where the subscriptions 1 and 2 denote the near and far points, respectively. 

{Since the thickness of the gas layer $z_{1/2}$ primarily depends on the Galactocentric distance, }
for a given LSR velocity, unknown parameters are $n_{01}$, $n_{02}$, $z_{01}$, and $z_{02}$, which we can estimate by fitting the model function to the data. A searching range of $z_{01}$ and $z_{02}$ is restricted to $|z|< 250$ pc as described in Paper II. 
To eliminate ill fitting results giving artificial structures, we smoothed the obtained distance-density curve using a spline function and clipping largely deviating points.  

\subsection{Vertical profile of HI and H$_2$ gases at tangential point}
Before applying the fitting method described in the previous subsection, we measured the thickness of the gas disk layer by analyzing the vertical ($z$-direction) profile of HI and H$_2$ gases at the tangential point, where the kinematic distance is uniquely determined for a given LSR velocity. Since the HI gas having an LSR velocity of the range $|V_{\rm t}|-\sigma \le |V_{\rm LSR}| \le |V_{\rm t}|$ contributes to a higher velocity component than the terminal velocity due the velocity dispersion $\sigma$, HI volume density was calculated by integrating the emission within the velocity range of $|V_{\rm t}|-\sigma \le |V_{\rm LSR}| \le \infty$ adopting $\sigma=5$ km s$^{-1}$ \citep{cle85}. Details are described in Papers I and II. 

{In the case of the outer Galaxy, the thickness was measured for all the points where vertical HI profile obtained. }

{The thickness $z_{1/2}$ was measured by fitting the model function to the observed vertical distributions. The measured FWHM ($2z_{1/2}$) of each gas layer is presented in section 4.4. }

\subsection{Regridding the obtained HI and H$_2$ distributions}
As described above, we obtained HI and H$_2$ density cubes in the form of $n_{\rm HI}(l, b, r)$ and $n_{\rm H_2}(l, b, r)$, from the line spectra cube $T_{\rm b}(l, b,V_{\rm LSR})$. Finally, these data cubes in the $(l,b,r)$ coordinates were transformed into $(x,y,z)$ coordinates explained in the section 1. 

For the outer Galaxy, the data were regridded by interpolating the nearest points. For the inner Galaxy, first the parameter sets of $z_0$ (midplane height) and $n_0$ (midplane density) were regridded by averaging ones surrounding a certain point with a weight function exponentially decreasing with FWHM of 0.48 kpc as described in Paper II. 

The foreground HI emission surrounding the Sun having LSR velocity of $V_{\rm LSR} \sim 0$ km s$^{-1}$ can contribute to density distribution around the solar circle $R=R_0$ where the gas giving LSR velocity of 0 km s$^{-1}$. To avoid this artificial effect, the data were regridded in the same way as the inner Galaxy for the region of $|V_{\rm LSR}|$ being less than $3\sigma$, instead of simple interpolation used for the outer Galaxy.

\subsection{Revision of HI map}
In this paper, HI distribution map is revised. The following points were revised: (1) searching range of $z_0$ is restricted to $-250$ -- $+250$ pc the same as Paper II, in calculating HI distribution in the inner Galaxy, (2) the parameter sets ($n_0$, $z_0$) obtained for the discrete points in the inner Galaxy were averaged with a Gaussian function whose FWHM was 0.48 kpc in order to regrid HI distribution data, and (3) flare component of the outer HI disk is not subtracted. Due to (1) and (2), artificial structures {seen in high-altitude of the inner Galaxy were able to be suppressed and the averaged surface density is not affected.}  
{Due to (3)}, the estimation of HI mass changed {by factor of 2}. {In Paper I, the most of the high-altitude emission was considered to be contributed from the local gas and this component was subtracted assuming that the local gas could be expressed by a linear function of the Galactic latitude $b$. Hence, the HI density at the midplane was also underestimated in this process. }
H$_2$ distribution map is {taken from the Paper II without any revision}. 


\section{New HI, H$_2$, and Total Gas Maps}
\subsection{Face-on Maps}
\label{faceonmap}
Figure \ref{HIH2maps} shows contour maps of resulting HI and H$_2$ column densities in the face-on view. 
The data range $|\theta| <30 \degree$ (i.e. behind the Galactic Center) was masked because data points are too sparse to make a map as described in Paper II. 
The global features such as spiral arms are consistent with previous works \citep{nak03,nak06a}. 

The H$_2$ gas is concentrated in the inner region ($\lesssim R_0$ kpc) and is hardly detectable in the outer region. On the other hand, the HI gas is distributed from the center to the outer region ($\gtrsim R_0$ kpc). 
The lowest contour of 1 M$_\odot$ pc$^{-2}$, which is often referred as a radius of gaseous disk \citep{bro94}, shows that the radius is 15--20 kpc and that the outskirt is not axissymmetric. The HI disk swells in the direction of $\theta = -45\degree$ if we assume a pure circular rotation. 
It is also shown that the disk is unnaturally asymmetry about $l=180\degree$. 
It is likely that the outermost disk does not circularly rotate with the outer disk. 

Next, we show comparisons of HI and H$_2$ in figure 2. The top panel of figure 2 is a false color image showing HI and H$_2$ in red and green, respectively. This figure shows obviously HI and H$_2$ gases are dominant in the outer and inner parts, respectively. The gas phase transition from HI to H$_2$ occurs within a narrow range of radius around $R_0$, which is referred as molecular front \citep{sof95, hon95}. 

The bottom panel of figure 2 is a total (HI plus H$_2$) gas map, which is presented for the first time by obtaining both HI and H$_2$ maps. {Figure 3 shows a contour map of the total gas map shown in figure 2. These maps are} advantageous to trace global structure such as spiral arms from the inner to the outer parts, since H$_2$ and HI {components preferentially traces the inner and outer arms,} respectively. Spiral arms traced here are described in the section 4.5. 

\begin{figure*}
  \begin{center}
    \rotatebox{-90}{\FigureFile(110mm,110mm){HI_gas_map.ps}}
    \rotatebox{-90}{\FigureFile(110mm,110mm){H2_gas_map.ps}}
  \end{center}
\caption{Face-on maps of (top) HI and (bottom) H$_2$ gases. Contour levels are 1.0, 1.4, 2.0, 2.8, 4.0, 5.6, 8.0, 11.3 M$_\odot$ pc$^{-2}$ for HI map and 0.25, 0.5, 1.0, 2.0, 4.0, 8.0, 16.0 M$_\odot$ pc$^{-2}$ for H$_2$. The area of $|\theta| < 30\degree$ is masked because the number of points, where HI and H$_2$ density are calculated, is too small to show the gas distribution.  \label{HIH2maps}}
\end{figure*}
\begin{figure*}
    \FigureFile(131mm,131mm){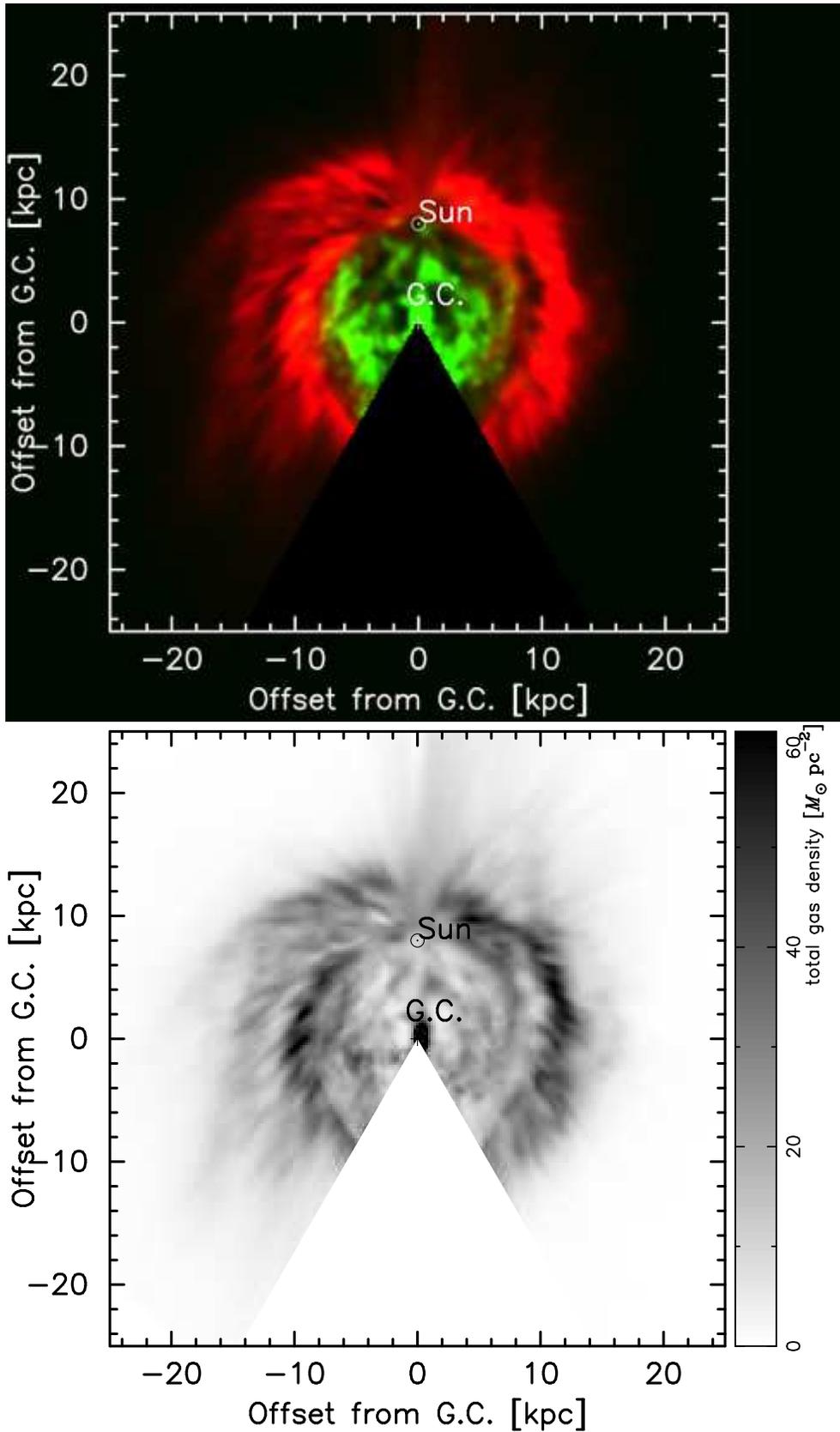}\\
    \rotatebox{-90}{\FigureFile(110mm,110mm){total_gas_map_nocont.ps}}
\caption{Top: Pseud color face-on map of HI (red) and H$_2$ (green) gases. Bottom: Grey scale face-on map of total gas distribution {combining HI and H$_2$ densities}. \label{totgas-map}}
\end{figure*}

\begin{figure*}
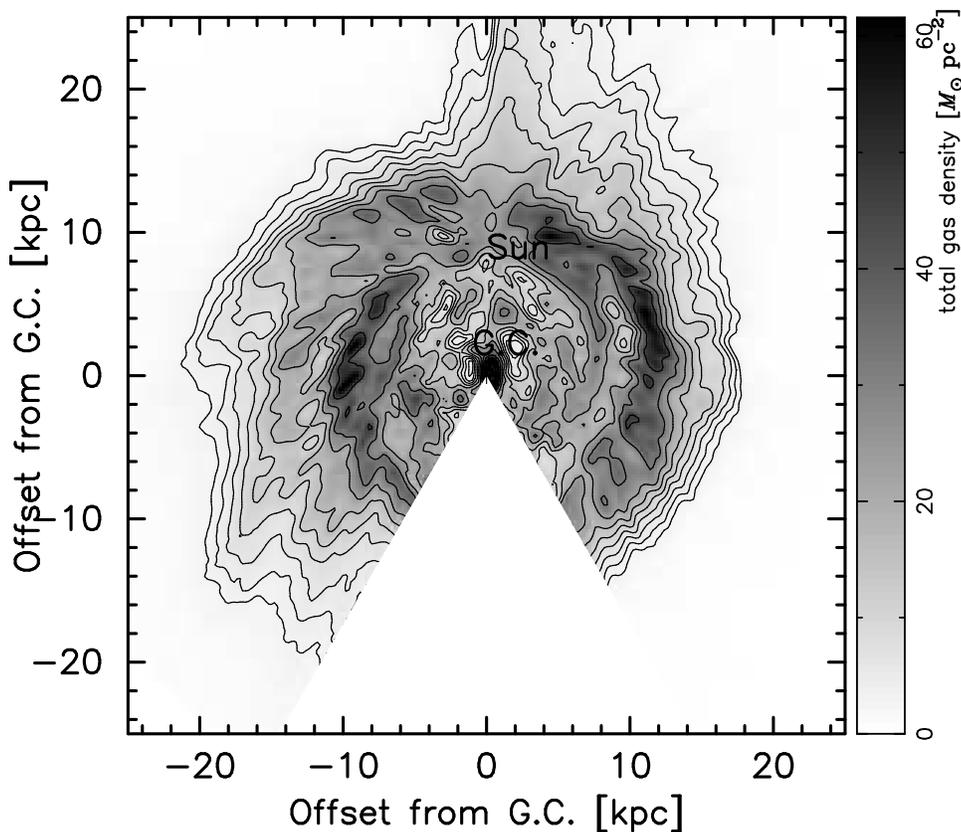

    \rotatebox{-90}{\FigureFile(110mm,110mm){total_gas_map2.ps}}
\caption{Contour map of total gas distribution shown in figure 2. Contour levels are 1.0, 1.4, 2.0, 2.8, 4.0, 5.6, 8.0, 11.3, 16.0 M$_\odot$ pc$^{-2}$.\label{totgas-map}}
\end{figure*}

\subsection{Radial distribution}
\begin{figure*}
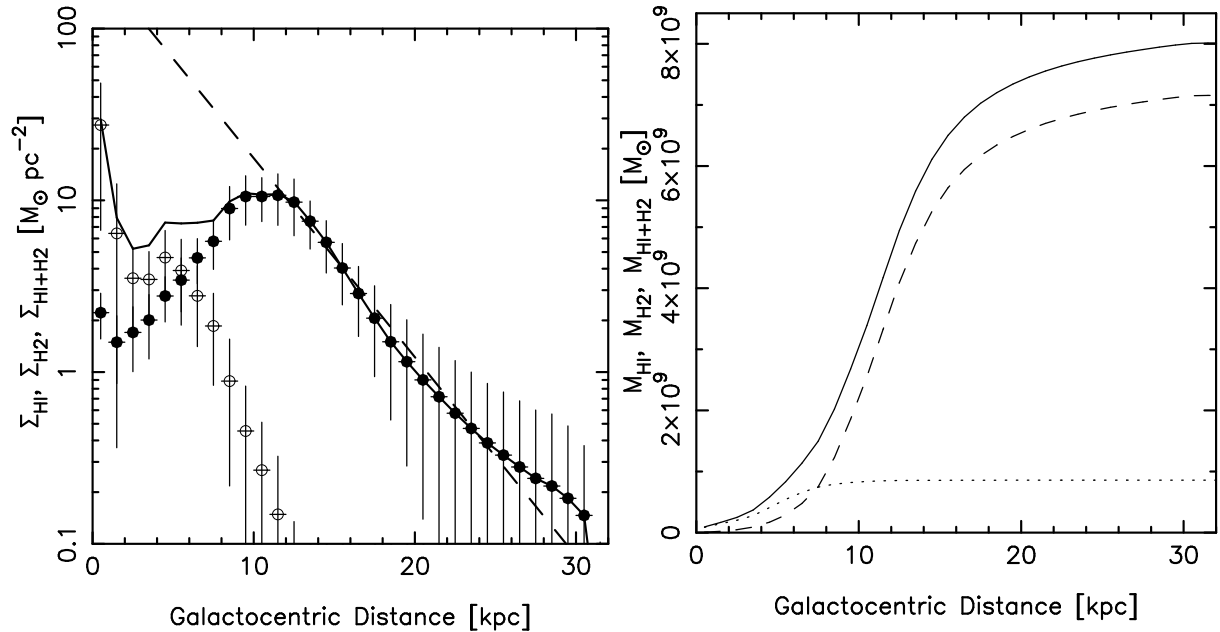

  \begin{center}
    \FigureFile(80mm,80mm){rgc_NHIH2.ps}
    \FigureFile(80mm,80mm){rgc_accum.ps}
  \end{center}
\caption{Left: Radial profiles of HI ({filled circles}), H$_2$ ({open circles}) and total gas surface densities (thick line). Dashed line is the fitted line taken from \citet{kal09}. Right: A function of the accumulated gas mass within each radius. {The dashed, dotted, and solid lines denote HI, H$_2$, and total gases, respectively.} H$_2$ and HI components reach the maximum value at the radii of 10 and 30 kpc, respectively.  \label{radpro-HICOtot}}
\end{figure*}

Left panel of figure \ref{radpro-HICOtot} shows radial distributions of the HI, H$_2$, and the total (HI plus H$_2$) gases, in {filled circles, open circles, and solid line}, respectively. This figure shows that the molecular gas is dominant in the central region and HI gas is dominant in the outer region. 
The total gas density attains its maximum at the center and second peak around $R=10$ kpc. Beyond $R=10$ kpc, it decreases with the Galactocentric distance and becomes lower than $1 M_\odot$ pc$^{-2}$ around $R=17$ kpc. The dashed line is exponential function expressed with $\Sigma (R)=\Sigma_0 e^{-(R-R_0)/R_h}$, where $\Sigma_0=30$ M$_\odot$ pc$^{-2}$ and $R_h=3.75$ kpc, adopted from \citet{kal09}. 

Right panel of figure \ref{radpro-HICOtot} shows the accumulated gas mass as a function of the Galactocentric distance. {HI and H$_2$ masses within the solar circle are alomost the same amount of $7.5\times 10^8$ M$_\odot$. This result is similar to \citet{blo86}, who estimated HI and H$_2$ masses to be $9.2\times 10^8$ M$_\odot$ and $9.4\times 10^8$ M$_\odot$, respectively,  by comparing high-energy gamma ray emission. Considering the difference of adopted Galactic constant, their averaged value $9.3\times 10^8$ M$_\odot$ is $9.3/7.5 \times (8/10)^2=0.8$ times smaller than our estimation. 
\citet{kal08} estimated the HI mass within the solar circle to be $1.8\times 10^9$ M$_\odot$. This discrepancy is thought to be caused by the difference of adopted models to derive HI distribution in the $z$ direction within the solar circle. In our paper, we took a model with sech$^2$ function as described in the section 3.3. On the other hand, \citet{kal08} took a model with an exponential function adopted in \citet{kal07}, which can reproduce HI distribution at high altitude. The HI surface density and HI mass, which we calculated, are thought to be underestimated by the amount of flared HI component as pointed out by \citet{loc84}. }

HI gas reaches the maximum value of $7.2\times 10^9$ M$_\odot$ at the radius of 30 kpc, which is consistent with \citet{kal09} considering the difference of adopted Galactic constant $R_0$. H$_2$ reaches the maximum value of $8.5\times 10^8$ M$_\odot$ around the radius of 12 kpc. The total gas mass reaches $8.0\times 10^9$ M$_\odot$ at the radius of 30 kpc. HI and H$_2$ gas components are 89 \% and 11 \% of the total gas mass, respectively.


\subsection{Vertical Cross Sections}
\begin{figure*}
    \FigureFile(85mm,85mm){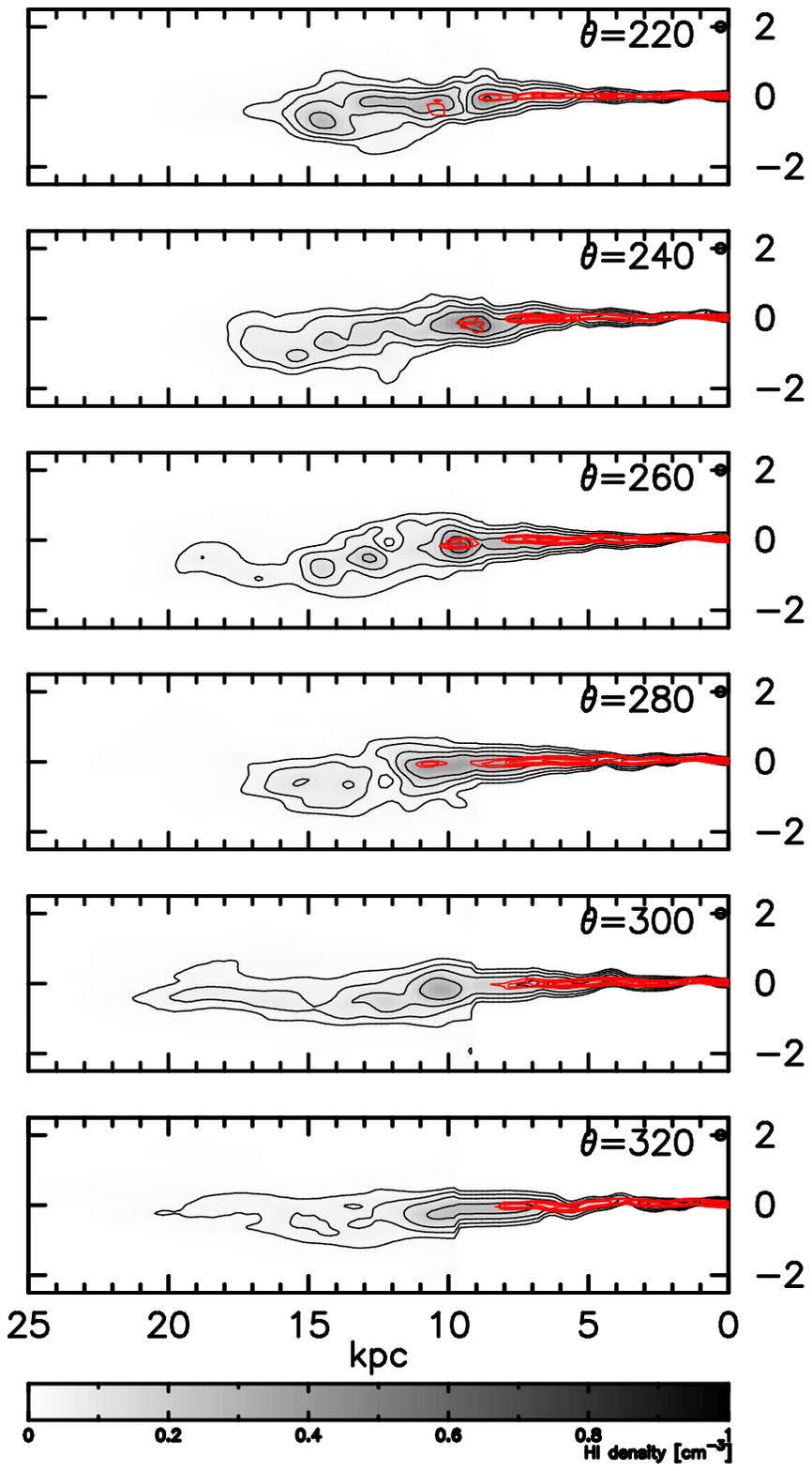}
    \FigureFile(85mm,85mm){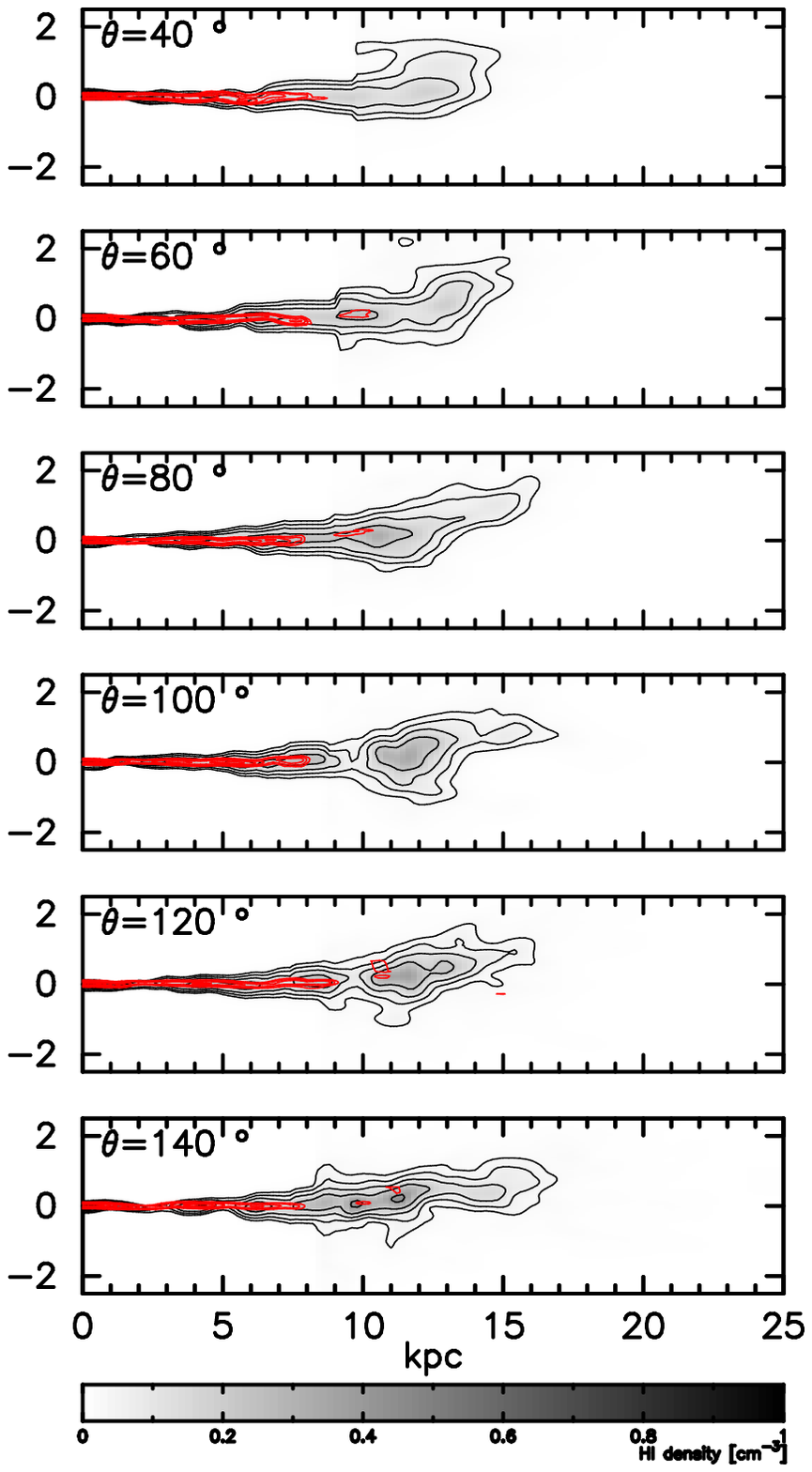}
\caption{Vertical slices of HI and H$_2$ gas maps. HI distribution is displayed with gray scale and black contours. H$_2$ map is presented with red contours. Contour levels are 0.025, 0.05, 0.10, 0.20, and 0.40 cm$^{-3}$. \label{verslice}}
\end{figure*}

Figure \ref{verslice} shows vertical cross sections of the HI and H$_2$ gas distributions. A gray scale and black contours denote the HI density. Red contours represent the H$_2$ gas density. 

The H$_2$ gas disk is thinner than the HI disk as external galaxies shows the same characteristics \citep{sco93}. The thickness of the HI gas disk increases with the Galactocentric distance as former works presented \citep{lev06,kal07}, while the thickness of the H$_2$ gas disk shows a smaller variation \citep{bro88,nak06a}. 

Figure \ref{verslice} shows that the peaks of the H$_2$ gas distribution roughly coincide with the HI gas peaks. However, midplane of H$_2$ disk is slightly shifted from the HI midplane in some parts as clearly seen in the outer disk of $\theta=120\degree$--$140\degree$. This can be interpreted that molecular gas fraction is not determined only with the pressure but UV radiation from star-forming regions changing the molecular fraction \citep{tan14} or that it is transient phase, in which molecular fraction leads to a equilibrium \citep{nak06b,inu15}. 


\subsection{Thickness of the Gas Disk}
\begin{figure}
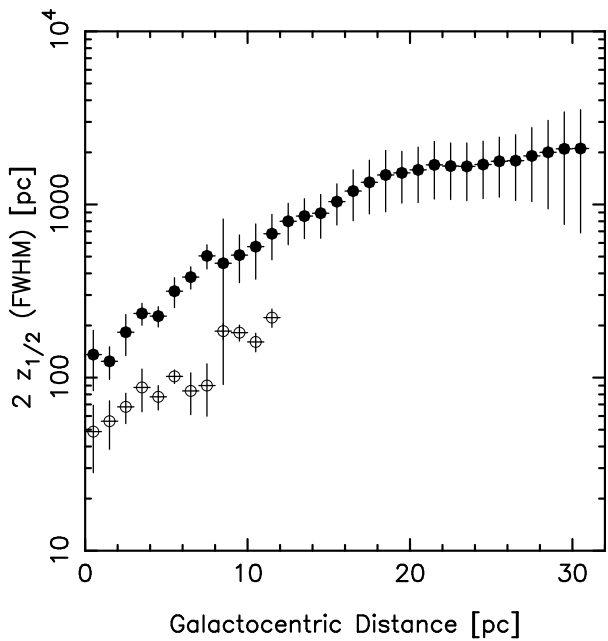

  \begin{center}
    \FigureFile(80mm,80mm){rgc_fwhm_log.ps}
  \end{center}
\caption{Thickness of the HI ({filled circle}) and H$_2$ ({open circle}) gas layers measured with $2z_{1/2}$ (FWHM) versus the Galactocentric distance.  \label{radpro-fwhm}}
\end{figure}

{The thickness of the gas layer is measured for each case of HI and H$_2$ as shown in figure \ref{radpro-fwhm}. 
The thickness of HI gas layer increases from 100 pc to 2 kpc with the radius, while the thickness of H$_2$ gas layer slowly increases from $\sim 50$ pc to $\sim 200$ pc. These results are consistent with the previous works (Papers I and II). The thin and small H$_2$ disk is embedded in the thicker and more extended HI gas disk as shown in a field galaxy \citep{cro01}.} 

\subsection{Spiral Arms}
\begin{figure*}
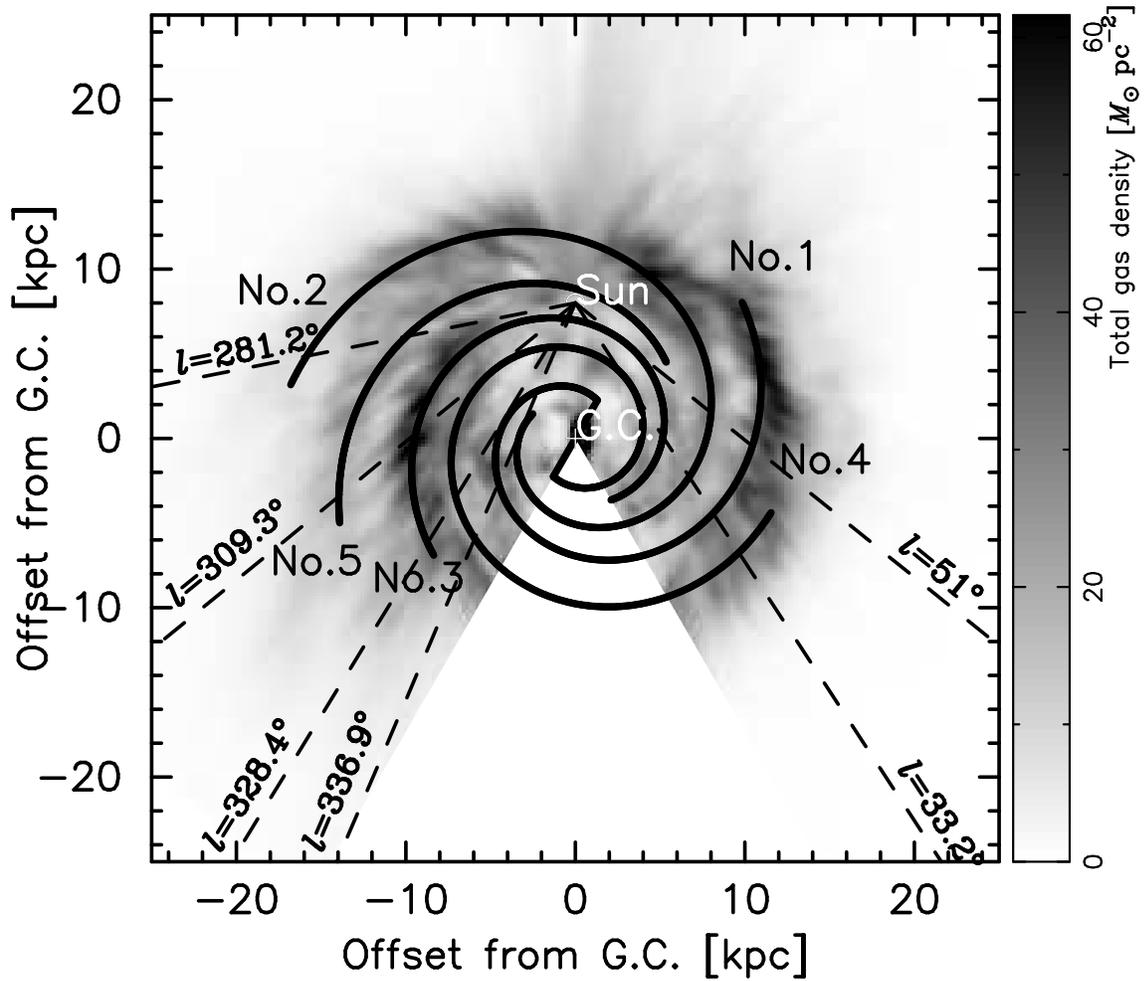

  \begin{center}
    \rotatebox{-90}{\FigureFile(130mm,130mm){total_gas_map_arms2.ps}}
  \end{center}
\caption{Schematic tracers of spiral arms superimposed on the total gas distribution map; No.1: Norma-Cygnus arm; No.2: Perseus arm; No.3: Sagittarius-Carina arm; No.4: Scutum-Crux arm; No.5: Orion (Local) arm. Dashed lines are tangents to the inner spiral arms. \label{totgas-arms}}
\end{figure*}

The total gas map shown in figure \ref{totgas-map} enable us to trace spiral arms from the inner to the outer Galaxy more clearly than ever, by taking advantages of the HI and H$_2$ maps. 
We can trace the well-known spiral arms such as the Norma-Cygnus, the Sagittarius-Carina, the Perseus, the Scutum-Crux, and the Orion arms from the inner to the outer Galaxy as shown in figure \ref{totgas-arms}, which shows schematic tracings of all these spiral arms. All the spiral arms can be traced as logarithmic spiral arms, which are {overlaid on the total gas map with thick lines. The pitch angles are in the range of $11\degree$--$15\degree$, which is the consistent with the mean value, $\timeform{13D.1} \pm \timeform{0D.6}$, of former works shown by \citet{val15}. The directions of tangents to individual arms are indicated with with broken lines. }
We describe properties of individual spiral arms in detail below.  

\begin{table*}
  \caption{Parameters of Spiral Arms}\label{parm_arms}
  \begin{center}
    \begin{tabular}{lcccc}
      \hline
      \hline
Name                 & Pitch angle & Beginning radius & Beginning angle & Ending angle \\
                     & ($\degree$)  & (kpc)            & ($\degree$)      & ($\degree$)   \\
      \hline
Norma-Cygnus (Outer) & 15          & 2.6              & $-210$            & 130          \\
Perseus              & 15          & 2.9              & $-120$            & 260          \\
Sagittarius-Carina   & 11          & 4.2              & $ 30$             & 310          \\
Scutum-Crux          & 11          & 2.6              & $-30$             & 430          \\
Orion (Local)        & 15          & 7.0              & $130$             & 290          \\
      \hline
     \hline
    \end{tabular}
  \end{center}
\end{table*}

%

\subsubsection{Norma-Cygnus Arm}
{The Norma arm is referred as the inner part of the spiral arm labelled with No.1. It is originally identified as a spiral arm found in the fourth quadrant in the inner Galaxy. The tangents to the Norma arm is found in the direction of $l=\timeform{328D.5} \pm \timeform{3D.5}$ \citep{val14}, which corresponds to the direction of the constellation Norma.  }

The Cygnus arm is referred as the outer part of the spiral arm labelled with No.1. {The Cygnus arm is also called the Outer arm \citep{val14}.} It is clearly traced in the face-on HI map as one of the most prominent arms in the outermost region, which arcs from $(R,\theta) \sim (9 \mbox{ kpc}, 50\arcdeg)$ to $(R,\theta)\sim (11 \mbox{ kpc}, 120\arcdeg)$ as described in Paper II. 
{Recent VLBI observation in H$_2$O masers line traced this spiral as a logarithmic spiral arm with pitch angle of $\timeform{13D.8} \pm \timeform{3D.3}$ based on the annual parallax \citep{hac15}}. 

As discussed in Paper II, the Norma and the Cygnus arms can be identified as the same spiral arm considering that their pitch angles should be $11\degree$ -- $15\degree$. Hence, we call this arm the Norma-Cygnus arm and labeled with No.1 as shown in figure \ref{totgas-arms}, where a schematic tracing of the Norma-Cygnus arm is presented by a logarithmic spiral arm starting at $(R, \theta)=(2.6 \mbox{ kpc}, -210\arcdeg)$ and ending at $\theta = 130\arcdeg$ with a pitch angle of $15\arcdeg$. Its origin is considered to coincide with ``near 3 kpc arm'' and to connect to the the near side of the Galactic bar \citep{val14}. {The parameters describing the arm is summarizes in Table \ref{parm_arms}. This picture is consistent with the recent panoramic view obtained with compilation of former works \citep{val05}. }

\subsubsection{Perseus Arm}
The Perseus arm labelled with No.2 is known as one of the most prominent arms \citep{geo76}. 
It is found in the range of $\theta = 30\arcdeg$ -- $230\arcdeg$, and is prominent in the outer Galaxy. 
The Perseus arm can be traced with a logarithmic spiral arm with a pitch angle of $15\arcdeg$. 
In figure \ref{totgas-arms}, a schematic tracing of the Perseus arm is shown by a logarithmic spiral arm starting at $(R, \theta)=(2.9 \mbox{ kpc}, -120\arcdeg)$ with a pitch angle of $15\arcdeg$ and ending at $\theta=260\arcdeg$. 

{\citet{val14} compiled the former works and concluded that the Perseus arm starts at the direction of $l=\timeform{336D.9}\pm \timeform{0D.7}0.7$, while \citet{chu09} concluded that the Perseus arm starts at the far edge of the central bar. 
If we assume that the Perseus arms start at the far edge of the central bar, the major radius of the central bar needs to be 5 kpc, considering the pitch angle and its distance measured around the Sun. However, the size of the bar and expected non-circular motion is inconsistent with former works. 
Our study supports the picture that the Perseus arm starts at $l=\timeform{336D.9}\pm \timeform{0D.7}$ but not from the either edge of the central bar, being concluded by \citet{val14}. }

\subsubsection{Sagittarius-Carina Arm}
The Sagittarius-Carina arm labelled with No.3 {in figure \ref{totgas-arms} is known as the closest spiral arm to the Sun. 
The tangents to the Sagittarius and Carina arms are found in the directions of $l=\timeform{48D.5}48.5\pm \timeform{2D.5}$ and $l=\timeform{284D.5} \pm \timeform{2D.5}$, which correspond to the directions of the Sagittarius and Carina constellations, respectively \citep{val14}. 

In this study,} the side of the Carina arm can be clearly traced in the HI and total gas maps and is well fitted by a logarithmic spiral arm with a pitch angle of $11\arcdeg$, though the side of the Sagittarius arm is less prominent as mentioned in Paper II. 
In figure \ref{totgas-arms} a schematic tracing of this arm is presented by a logarithmic spiral arm starting at $(R, \theta)=(4.2 \mbox{ kpc}, 30\arcdeg)$ with a pitch angle of $11\degree$ and ending at $\theta=310\arcdeg$.

\subsubsection{Scutum-Crux Arm}
The Scutum-Crux arm labelled with No.4 is found between the Norma-Cygnus and the Sagittarius-Carina arms as indicated in figure \ref{totgas-arms}. {Similarly to the Sagittarius-Carina arm, the tangent to the Scutum and Crux arms are found in the directions of $l=\timeform{33D.2} \pm \timeform{17D.2}$ and $l=\timeform{309D.3} \pm \timeform{0D.9}$ \citep{val14}, which correspond to the directions of the Scutum and Crux constellations, respectively. The Crux arm often referred as Centaurus arm and so it is often called {Scutum}-Centaurus \citep{chu09}. }

In the first quadrant, {the separation between the Scutum-Crux and Sagittarius-Carina arms gets small and they seem form a ring-like structure, which was reported to exist at the Galactocentric radius of $R_0/2$ \citep{cle88}.}  

This arm can be traced with a logarithmic spiral arm with a pitch angle of $11\arcdeg$. 
{Considering its pitch angle, it is thought to connect to the distant molecular arm found by \citet{dam11} as they concluded. }
In figure \ref{totgas-arms} a schematic tracing of the Scutum-Crux arm is shown by a logarithmic spiral arm starting at $(R, \theta)=(2.6 \mbox{ kpc}, -30\arcdeg)$ and ending at $\theta=430\arcdeg$ with a pitch angle of $11\arcdeg$.

\subsubsection{Orion Arm (Local Arm)}
Between the Sagittarius--Carina and the Perseus arms, there exists a less prominent but clear spiral arm, where the Sun is located, as shown in figure \ref{totgas-arms}. This spiral arm is the referred as the Orion arm or the Local arm \citep{bob14}.    
In figure \ref{totgas-arms}, a schematic tracing of the Local arm is presented by a logarithmic spiral arm starting at $(R, \theta)=(7.0 \mbox{ kpc}, 130\arcdeg)$ and ending at $\theta = 290\arcdeg$ with a pitch angle of $15\degree$.

{So far, there are two possibilities on the structure of the Orion arm; (1) it is a bridge-like structure connecting the Sagittarius-Carina and Perseus arms \citep{wea70,chu09}, or (2) normal spiral arm with similar pitch angle as the others. Our result supports the latter picture, which is is consistent with recent VLBI result concluded by \citet{xu13} and \citet{bob14}, who measured annual parallaxes of 23--25 high-mass star formation regions with VLBA to show that the Orion arm is traced with a logarithmic spiral arm with the pitch angle of $\timeform{12D.9}\pm \timeform{2D.9}$ -- $\timeform{13D.8} \pm \timeform{3D.3}$. }

\subsection{Molecular Fraction}
\begin{figure}
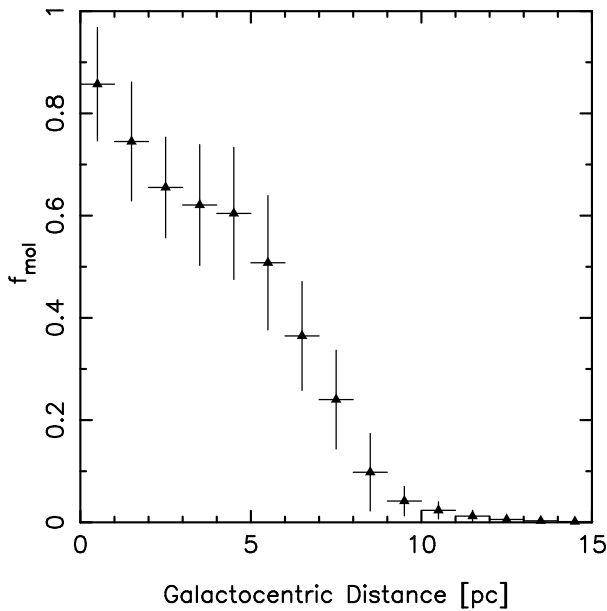

  \begin{center}
    \rotatebox{0}{\FigureFile(80mm,80mm){rgc_fmol.ps}}
  \end{center}
\caption{Radial variation of the molecular fraction ($f_{\rm mol}$) obtained by azimuthally averaging two dimensional $f_{\rm mol}$ map. \label{r-fmol}}
\end{figure}

We here define a parameter $f_{\rm mol}$ to quantitatively investigate the physical properties of the ISM as follows, 
\begin{equation}
f_{\rm mol}={\Sigma_{{\rm H}_2} \over \Sigma_{{\rm HI}} + \Sigma_{{\rm H}_2}}, 
\end{equation}  
where $\Sigma_{{\rm HI}}$ and $\Sigma_{{\rm H}_2}$ are surface densities of HI and H$_2$ gases, respectively. 
Figure \ref{r-fmol} shows $f_{\rm mol}$ plotted versus Galactocentric distance. The molecular fraction $f_{\rm mol}$ is {above 0.5 in the Galactocentric range of ($R<5$ kpc)} , which implies that H$_2$ gas is dominant {in this range}. {The molecular fraction $f_{\rm mol}$ rapidly decreses beyond $R=5$ kpc and }
gets less than 0.1 in the outer Galaxy ($R>8$ kpc), which implies that HI gas is dominant {in the outer disk}. 

\begin{figure}
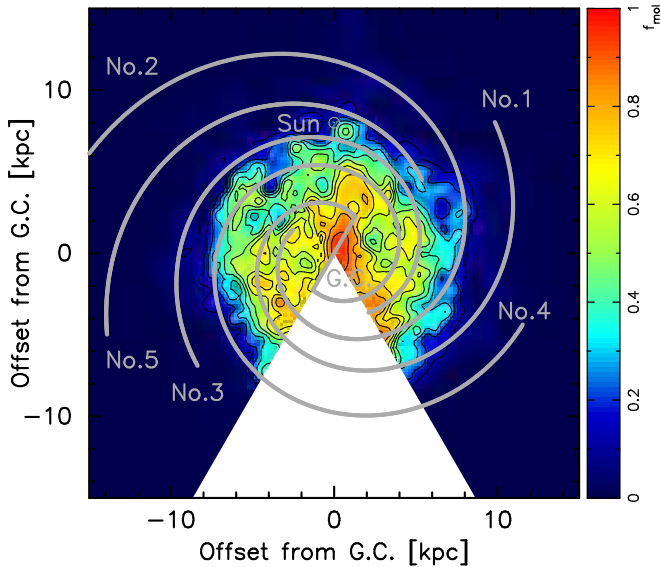

  \begin{center}
    \rotatebox{-90}{\FigureFile(75mm,75mm){fmol_map.ps}}
  \end{center}
\caption{Two-dimensional molecular fraction ($f_{\rm mol}$) map calculated with HI and H$_2$ surface densities. The molecular faction $f_{\rm mol}$ is defined as $f_{\rm mol}={\Sigma_{{\rm H}_2} /(\Sigma_{{\rm HI}} + \Sigma_{{\rm H}_2})}$. The schematic tracers of spiral arms are indicated with gray lines.  \label{2Dfmol-map}}
\end{figure}

Figure \ref{2Dfmol-map} shows a map of two-dimensional $f_{\rm mol}$ distribution. In figure \ref{2Dfmol-map} we superimposed schematic tracings of spiral arms with gray lines. These spiral arms are the same as those described in Section \ref{faceonmap}. 
In addition to the global variation, local variations of the molecular fraction are found. The molecular fraction $f_{\rm mol}$ tends to increase along the spiral arms rather than the inter-arm regions. The molecular fraction $f_{\rm mol}$ varies by 0.1 -- 0.2 between an arm and an inter-arm. This tendency is also found in external galaxies \citet{kun95}. 

\section{Possible Uncertainties}
\begin{figure}
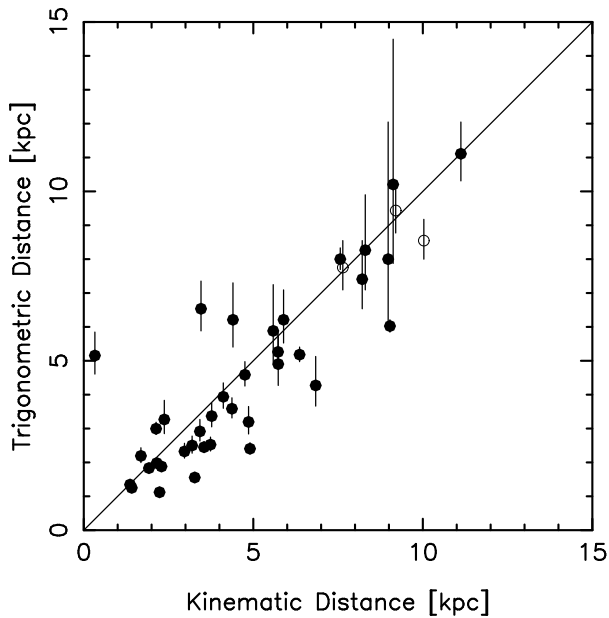

  \begin{center}
    \FigureFile(80mm,80mm){kindist_pardist.ps}
  \end{center}
\caption{Comparison of kinematic distance used in this paper and trigonometric distance obtained with VLBI observation \citep{rei14}. Open and filled circles indicate sources within the Galactocenric distance of 3 kpc and the others, respectively. \label{kindist_pardist}}
\end{figure}

{We adopted several assumptions for the simplicity to obtain the gaseous maps. In this section, we note possible systematic errors caused by such assumptions. 

\noindent{1. Kinematic Distance}\\
The kinematic distance, which we adopted as a distance estimator, is based on the assumption that the Galactic rotation is purely circular. However, it is known that non-circular motion exists and affects the kinematic distance. 

There have been a lot of efforts to the measurements of annual parallaxes with VLBI, which have shown that typical error is within a few kpc \citep{hac09,sak12,nak15}. {Figure \ref{kindist_pardist} shows comparison of VLBI trigonometric distance taken from \citet{rei14} and kinematic distance used in this paper. Open and filled circles indicate sources within the Galactocenric distance of 3 kpc and the others, respectively. If the both values are the same, the data are plotted on the linear line.  

The standard deviation between trigonometric and kinemactic distances is calculated to be 2.3 kpc, which is a typical error in heliocentric distance of the maps shown in this paper. This value is consistent with results from the recent an N-body+hydrodynamical simulation shown by 
\citet{bab09}, who concluded that kinematic distance had error of 2--3 kpc. 

Figure \ref{kindist_pardist} shows that there is no big difference between filled and open circles though the non-circular motion is significant near the Galactic center (e.g., Sawada et al. 2004).  Therefore, the typical error of the kinemetic distance of the sources within the Galactocentric distance of 3 kpc is of the same order of the other sources located in the outer region.}

\noindent{2. Galactic Constant}\\
The Galactic constants ($R_0$, $V_0$) change the kinematic distance, according to the equation (1). Since the Galactocentric distance of a gas cloud having LSR velocity $V_{\rm LSR}$ in the direction of galactic longitude $l$ is calculated to be $R=R_0/({V_{\rm LSR}\over V(R)\sin{l}} + {V_0\over V(R)})$, the solar Galactocentric distance $R_0$ essentially scales the size but not the shape of the gaseous disk. The surface density changes by square of $R_0$. 

Though the solar rotational velocity $V_0$ can change the kinematic distance, it is order of ${V_0/V(R)}$, which is typically less than 25 \% as is estimated from fluctuations of the rotation curve \citep{sof09}. 

\noindent{3. Fitting Method}\\
In order to divide the mixture emission from near and far points of the inner Galaxy, we utilized the fitting method as explained in section 3.3. However, it is not necessarily perfect because the actual HI vertical distribution cannot expressed with a single function \citep{loc84}. A difference of midplane density between the data and model is typically less than 20 -- 30\% as mentioned in Paper II.

\noindent{4. Optical Thickness} \\
In this study, we assume that the HI line is optically thin to measure the HI column density for the simplicity. However, recent studies point out that HI mass is underestimated by factor 2--2.5 when HI line is assumed to be optically thin \citep{fuk14,fuk15}. 
}

\section{Summary}
In this paper, we showed three dimensional HI and H$_2$ maps of the Milky Way Galaxy using the latest HI and CO survey data and rotation curves based on the kinematic distance. The HI map has been revised with the improved analysis from the Paper I. 

{Comparing the HI and H$_2$ maps, it was shown that 
the} H$_2$ gas is distributed mainly inside the solar circle while HI distribution is extended to the large Galactocentric distance of 15--20 kpc and its outskirt is asymmetric about the Galactic center. 
The radial distribution of total gas density generally decreases with the Galactocentric distance peaking at the center while it is roughly constant between 5 and 10 kpc. The total gas mass is $8\times 10^9$ M$_\odot$, 89 \% and 11\% of which are HI and H$_2$ gas masses, respectively. 
The H$_2$ gas layer is embedded in the thicker and more extended HI layer. The HI layer is largely bent at the Galactocentric distance of 12 kpc as well as is corrugated in the inner Galaxy. 

We show the total gas density map of the Milky Way Galaxy for the first time by summing up the HI and H$_2$ gas density map.  
It is more advantageous to trace spiral arms from the center to the outer Galaxy than the individual HI and H$_2$ maps. 
We could easily trace five logarithmic spiral arms in the total gas map such as the Norma-Cygnus, the Perseus, the Sagittarius-Carina, the Scutum-Crux, and the Orion arms. The Norma and the Cygnus arms are identified as the same spiral arm, considering their pitch angles of $11\degree$ -- $15\degree$. 

The fraction of molecular component to the total gas decreases with the Galactocentric distance and drops steeply in the small radius range of 5 -- 8 kpc. The two dimensional molecular fraction map shows that it is also locally enhanced on the spiral arms. 

\noindent{\bf Acknowledgement}
We would like to thank the referee for carefully reading the manuscript. This work was supported by JSPS KAKENHI Grant Number 26800104.

\end{document}